\documentclass[conference]{IEEEtran}
\usepackage[utf8]{inputenc}
\usepackage[T1]{fontenc} 
\IEEEoverridecommandlockouts
\usepackage{cite}
\usepackage{amsmath,amssymb,amsfonts}
\usepackage{algorithmic}
\usepackage{graphicx}
\usepackage{textcomp}
\usepackage{xcolor}
\usepackage{subcaption}
\usepackage{bm}
\usepackage{booktabs}

\def\BibTeX{{\rm B\kern-.05em{\sc i\kern-.025em b}\kern-.08em
    T\kern-.1667em\lower.7ex\hbox{E}\kern-.125emX}}

\newcommand{\noofEEG}{N}
\newcommand{\noofchn}{M}
\newcommand{\chnidx}{m}
\newcommand{\nooftimepoints}{T}
\newcommand{\timeidx}{t}
\newcommand{\eegimage}{\bm{X}}
\newcommand{\eegimageperelm}{X}
\newcommand{\latentfeat}{\bm{Z}}
\newcommand{\latentdim}{D}
\newcommand{\maskratio}{r}
\newcommand{\masklength}{l_m}
\newcommand{\unmasklength}{l_u}
\newcommand{\maskedset}{\mathcal{M}}

\begin{document}

\title{Unsupervised Multivariate Time-Series Transformers for Seizure Identification on EEG\\
\thanks{This is a pre-print version of the article: I.\ Yıldız Potter, G.\ Zerveas, C.\ Eickhoff, D.\ Duncan, ``Unsupervised Multivariate Time-Series Transformers for Seizure Identification on EEG'', IEEE Conference on Machine Learning and Applications (ICMLA) 2022, DOI 10.1109/ICMLA55696.2022.00208.
Work was performed at University of Southern California and supported by the National Institutes of Health (NIH) National Institute of Neurological Disorders and Stroke (NINDS) grant R01NS111744. Code is publicly available at \texttt{https://github.com/ilkyyldz95/EEG\_MVTS}.}
}

\author{\IEEEauthorblockN{\.{I}lkay Y{\i}ld{\i}z Potter}
\IEEEauthorblockA{\textit{BioSensics LLC} \\
Newton, MA, USA \\
ilkay.yildiz@biosensics.com}
\and
\IEEEauthorblockN{George Zerveas}
\IEEEauthorblockA{\textit{Dept. of Computer Science} \\
\textit{Brown University}\\
Providence, RI, USA \\
george\_zerveas@brown.edu}
\and
\IEEEauthorblockN{Carsten Eickhoff}
\IEEEauthorblockA{\textit{Dept. of Computer Science} \\
\textit{Brown University}\\
Providence, RI, USA \\
carsten@brown.edu}
\and
\IEEEauthorblockN{Dominique Duncan}
\IEEEauthorblockA{\textit{Keck School of Medicine} \\
\textit{University of Southern California}\\
Los Angeles, CA, USA \\
Dominique.Duncan@loni.usc.edu}
}

\maketitle

\begin{abstract}
Epilepsy is one of the most common neurological disorders, typically observed via seizure episodes. Epileptic seizures are commonly monitored through electroencephalogram (EEG) recordings due to their routine and low expense collection. The stochastic nature of EEG makes seizure identification via manual inspections performed by highly-trained experts a tedious endeavor, motivating the use of automated identification. The literature on automated identification focuses mostly on supervised learning methods requiring expert labels of EEG segments that contain seizures, which are difficult to obtain. Motivated by these observations, we pose seizure identification as an \emph{unsupervised anomaly detection} problem.
To this end, we employ the first unsupervised transformer-based model for seizure identification on raw EEG. We train an autoencoder involving a transformer encoder via an unsupervised loss function, incorporating a novel masking strategy uniquely designed for multivariate time-series data such as EEG. Training employs EEG recordings that do not contain any seizures, while seizures are identified with respect to reconstruction errors at inference time. We evaluate our method on three publicly available benchmark EEG datasets for distinguishing seizure vs.~non-seizure windows. Our method leads to significantly \emph{better seizure identification performance than supervised learning counterparts}, by up to $16\%$ recall, $9\%$ accuracy, and $9\%$ Area under the Receiver Operating Characteristics Curve (AUC), establishing particular benefits on highly imbalanced data.  Through accurate seizure identification, our method could facilitate widely accessible and early detection of epilepsy development, without needing expensive label collection or manual feature extraction.
\end{abstract}


\section{Introduction}
\label{sec:intro}

Epilepsy is one of the most common neurological disorders, affecting over 70 million people worldwide \cite{thijs2019epilepsy}. Epilepsy patients typically suffer from seizures, involving uncontrolled jerking movements or momentary losses of awareness due to abnormal excessive or synchronous activities in the brain \cite{vespa2019epilepsy}. The degraded quality of life for patients strongly motivates early seizure identification, as early seizures have been shown to be prognostic markers for later epileptogenic development. Successful identification of early seizures can initiate antiepileptogenic intervention and therapies that can remarkably improve the quality of life for patients and their caregivers. To this end, electroencephalogram (EEG) recordings received particular attention for seizure identification \cite{staba2014electrophysiological}, due to their routine and low expense collection compared to, e.g., neuroimaging. Seizures on EEG are defined as generalized spike-wave discharges at three per second or faster, and clearly evolving discharges of any type that reach a frequency of four per second or faster.

Despite their volume and rich information content, EEG recordings are known to contain many artifacts  due to movement, physiological activity such as perspiration, and measurement hardware \cite{deng2018transductive, saba2021unsupervised}. The stochastic nature of clinically-acquired EEG makes seizure identification via manual inspection laborious and difficult, leading to significant variability across clinical labels of different experts \cite{zhang2016lmd}. 
This challenge motivated the recent literature to focus on automated identification of epileptic seizures on EEG as a promising complement to manual inspection. 
The literature on automated EEG seizure identification is extensive (c.f.~Section \ref{sec:related}), focusing mostly on \emph{supervised} machine learning methods using both manual feature extraction \cite{amin2020novel, ramos2020feature, 9441235, mehla2021efficient}, as well as deep neural networks (DNNs) without manual feature extraction \cite{zhang2020adversarial, li2020epileptic, eldele2021time}. 

Despite their success, supervised methods require expert labels indicating EEG segments that contain seizures, while obtaining large and consistently-labeled EEG datasets is unfavourable due to the stochastic nature of EEG \cite{zhang2016lmd}. Difficulty of label collection also leads to severely imbalanced EEG datasets, in which the number of non-seizure recordings significantly exceeds the number of seizure recordings; this poses a further challenge for supervised learning that is prone to overfitting towards dominant class predictions \cite{islah2020machine}.   

Unsupervised machine learning methods, which do not rely on labeled data have not yet been widely explored. A few methods employed traditional shallow models for unsupervised seizure identification on both raw EEG \cite{Chakrabarti2017}, as well as spatio-temporal features extracted from EEG \cite{belhadj2016whole, birjandtalab2016unsupervised, charupanit2020detection}.  
To the best of our knowledge, unsupervised DNN methods for EEG seizure identification have been limited to a couple of recent works, requiring feature extraction prior to training \cite{you2020unsupervised} or employing convolutional DNN architectures that are not tailored for multivariate time-series data such as EEG \cite{YILDIZ2021106604}.

We propose a \emph{fully-unsupervised} deep learning approach that can identify seizures on \emph{raw} EEG recordings. To this end, we make the following contributions:
\begin{itemize}
    \item We employ the first unsupervised transformer-based model for seizure identification on raw EEG, inspired by recent advances in multivariate time-series analysis \cite{zerveas2021transformer}. 
    \item We pose seizure identification as an anomaly detection problem. To this end, we train an autoencoder involving a transformer encoder via an unsupervised loss function, incorporating a novel masking strategy uniquely designed for modeling multivariate time-series  data such as EEG. As training employs EEG recordings that \emph{do not} contain seizures, seizures are identified via mean reconstruction errors at inference time.
    \item We extensively validate the seizure identification performance of our method on three publicly available benchmark EEG datasets. Our method can successfully distinguish between non-seizure vs.~seizure windows, with up to \emph{0.94 Area under the Receiver Operating Characteristics Curve (AUC)}. Moreover, our unsupervised anomaly detection approach leads to significantly \emph{better seizure identification performance than the supervised learning counterparts}, by up to $16\%$ recall, $9\%$ accuracy, and $9\%$ AUC, establishing a particular benefit for learning from highly imbalanced data.
\end{itemize}

\section{Related Work}
\label{sec:related}

The literature on automated seizure identification on EEG is vast; we refer the reader to the review by \cite{boonyakitanont2020review} for more details. 
A significant body of works focus on extracting spatio-temporal features from EEG via, e.g., wavelet transformations \cite{amin2020novel, radman2020multi}, local mean decomposition \cite{zhang2016lmd}, Fourier transformations \cite{ramos2020feature, mehla2021efficient}, and power spectra \cite{9441235}. 
Extracted features are used to train \emph{supervised} machine learning methods, including support vector machines and neural networks, to identify whether a given EEG contains a seizure in a binary classification setting. 

Deep neural network (DNN)-based supervised seizure identification methods have lately dominated the literature \cite{zhao2020novel} and obviated the need for manual feature extraction. DNN methods further improved in combination with recurrent neural networks to aid time-series modeling \cite{chakrabarti2021channel}, adversarial training to generalize identification across patients \cite{zhang2020adversarial}, autoencoder-based feature extraction \cite{sun2019unsupervised}, and attention mechanisms to improve predictions and interpretability \cite{li2020epileptic}.

In recent years, self-attention modules have become an integral part of DNN methods employed in machine vision \cite{dosovitskiy2020image}, natural language processing \cite{devlin2019bert}, and time-series modeling \cite{zerveas2021transformer}; the resulting DNN architectures are termed as \emph{transformers}. Transformer architectures have been very recently applied for various identification tasks on EEG, including, e.g., sleep-stage classification , human-computer interface-based action recognition, and seizure identification \cite{eldele2021time, kostas2021bendr}. These methods employ unsupervised pre-training prior to supervised training on ground-truth expert labels pertaining to the identification task. The unsupervised pre-training objective involves different augmentations of the same EEG segment and trains the transformer by maximizing (minimizing) the similarity of different augmentations (segments).

All in all, the literature on automated seizure identification often focuses on supervised machine learning methods. Despite their success, these methods require expert labels indicating EEG segments that contain seizures, which are difficult to obtain due to the stochastic nature of EEG \cite{zhang2016lmd}.
Meanwhile, unsupervised machine learning methods that do not rely on labeled data have not yet been widely explored. A few methods applied shallow models for unsupervised seizure identification, including K-means, hierarchical clustering, and Gaussian mixture models, on both raw EEG \cite{Chakrabarti2017}, as well as spatio-temporal features extracted from EEG \cite{belhadj2016whole, birjandtalab2016unsupervised}.  

Recently, a couple of unsupervised DNN methods for seizure identification on EEG have been proposed. You et al.~(2020) preprocess EEGs to extract time-frequency spectrogram images and train a generative adversarial network (GAN) \cite{you2020unsupervised} on the spectrograms that do not contain seizures. For each spectrogram at testing time, they have to search for the latent GAN input that leads to the smallest loss value and use the corresponding generated spectrogram for seizure identification. As training involves non-seizure activity, test spectrograms that significantly differ from the spectrograms generated by the GAN are successfully identified to contain seizures. Yıldız et al.~(2022) train a convolutional variational autoencoder (VAE) over raw EEG, employing an objective tailored for suppressing EEG artifacts. Unlike You et al.~(2020), they identify seizures with respect to the reconstruction errors at inference time. 

We differ from the existing works by applying the first \emph{fully-unsupervised transformer-based} model on \emph{raw} EEG. Our architecture and training objective are particularly designed for multivariate time-series analysis and do not require a sophisticated minimax optimization such as GAN training. The fundamental benefit of a transformer encoder over other DNN architectures is that self-attention can selectively highlight important input features and sequence segments, without relying on sequence-aligned convolutions or slow recurrent modules \cite{vaswani2017attention}; we also experimentally demonstrate this advantage against unsupervised VAE-based seizure identification in Section \ref{sec:results}.

\section{Problem Formulation}
\label{sec:method}

We consider a dataset of $\noofEEG$ EEG recordings, each collected from $\noofchn$ electrode channels and consisting of $\nooftimepoints$ time points. Formally, we denote each EEG recording by $\eegimage^{(i)} \in \mathbb{R} ^ {\nooftimepoints \times \noofchn}$, for $i \in [1,\ldots,\noofEEG]$. Our aim is to design an unsupervised method that does not rely on ground-truth expert labels during learning and can identify the existence of seizures in a given EEG recording. To this end, we employ an autoencoder architecture involving a transformer network encoder that is uniquely designed for multivariate time-series data \cite{zerveas2021transformer}, such as EEG. 
We note that our method naturally generalizes to EEG recordings comprising different numbers of time points and channels (see our preprocessing setup in Section \ref{sec:preprocessing}).

\begin{figure*}[!t]
\centering
\includegraphics[width=0.95\linewidth]{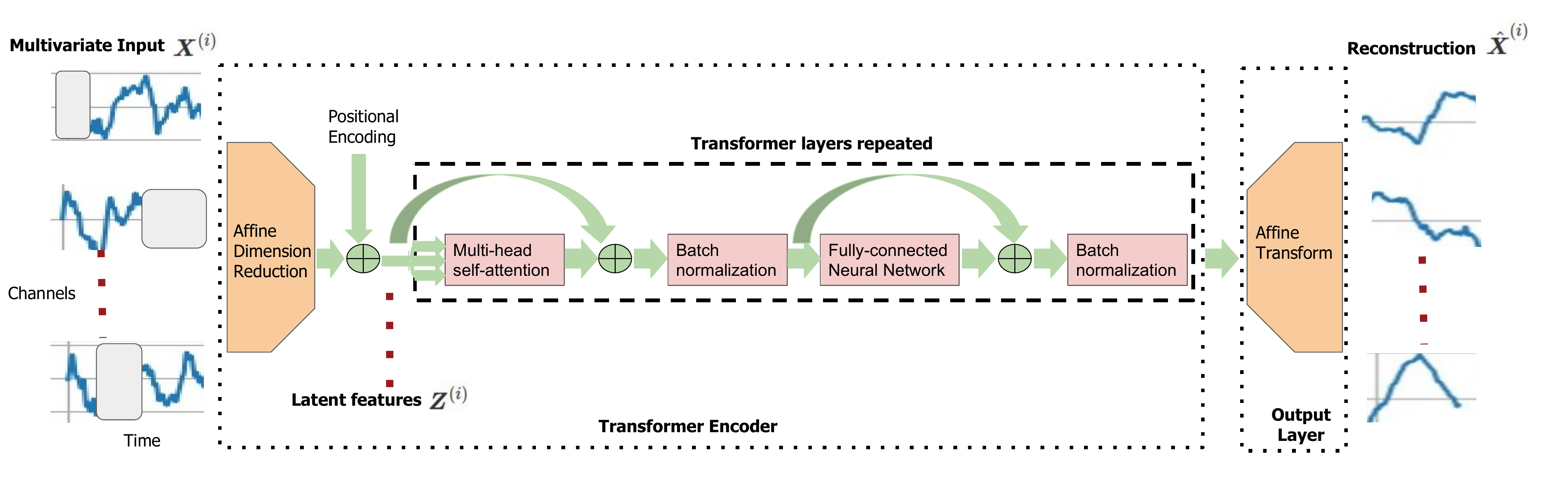}
\vspace{-0.75em}
\caption{Our autoencoder architecture. The transformer encoder network receives a recording $\eegimage^{(i)}$, and extracts latent features $\latentfeat^{(i)}$. The output layer applies an affine transformation on $\latentfeat^{(i)}$ to reconstruct the recording as $\hat \eegimage^{(i)} \in \mathbb{R} ^ {\nooftimepoints \times \noofchn}$. During training, a proportion of each channel is masked by setting the input values at masked time points (shaded in gray) to 0.}
\vspace{-0.5em}
\label{fig:architecture}
\end{figure*}

\subsection{Multivariate Time-Series Transformer}
\label{sec:architecture}
Our autoencoder architecture is based on a transformer encoder and is depicted in Figure \ref{fig:architecture}: the model learns to extract and transform latent features from a given EEG recording in order to reconstruct the stochastically-masked input \cite{zerveas2021transformer}. Formally, the transformer encoder network receives a recording $\eegimage^{(i)} \in \mathbb{R} ^ {\nooftimepoints \times \noofchn}$, $i \in [1,\ldots,\noofEEG]$, and extracts latent features $\latentfeat^{(i)} \in \mathbb{R}^{\nooftimepoints \times \latentdim}$. The output layer applies an affine transformation on $\latentfeat^{(i)}$ to reconstruct the recording as $\hat \eegimage^{(i)} \in \mathbb{R} ^ {\nooftimepoints \times \noofchn}$.

\subsubsection{Transformer Encoder} Transformer encoder operations begin with projecting a recording $\eegimage^{(i)}$ from $\noofchn$ dimensions to $\latentdim$ dimensions via a trainable affine transformation $\bm{P} \in \mathbb{R}^{\noofchn \times \latentdim}$. To preserve the ordering information of the input sequence, a fully-trainable positional encoding $\bm{E} \in \mathbb{R}^{\nooftimepoints \times \latentdim}$ is added for each input. The resulting latent features extracted from each recording are thus: $\latentfeat^{(i)} = \eegimage^{(i)}\bm{P} + \bm{E}$. 

Dimensional projection and positional encoding are followed by the successive application of several transformer layers. Each transformer layer consists of a multi-headed self-attention (MSA) module, a stochastic dropout operation $\tilde{d}$~\cite{dropout}, batch normalization (Norm) \cite{ioffe2015batch}, and a fully-connected network (FCN) consisting of two linear layers separated by a GELU \cite{gelu} activation, a non-linearity designed to be used in combination with dropout and batch normalization. Formally, latent features are updated by each transformer layer via:
\begin{align}
    \latentfeat^{(i)} \leftarrow \text{Norm}\left (\mathop{\tilde{d}}\left (\text{MSA}(\latentfeat^{(i)})\right ) + \latentfeat^{(i)} \right), \nonumber\\
    \latentfeat^{(i)} \leftarrow \text{Norm}\left(\mathop{\tilde{d}}\left (\text{FCN}(\latentfeat^{(i)})\right) + \latentfeat^{(i)}\right).
\end{align}
The summation of each latent feature with its transformation is a \emph{skip connection} that aids generalization \cite{he2016deep}, along with batch normalization that has been shown improvement against layer normalization for multivariate time-series analysis \cite{zerveas2021transformer}.  

\subsubsection{Multi-headed Self-attention Module}
An MSA module is designed to assign selective importance to latent features extracted for each time point by the preceding layers of the encoder \cite{vaswani2017attention}. Particularly, MSA contains trainable parameters that capture the similarity between input features at different time points via their query, key, and value representations. Multiple attention heads enable adaptations to long-term dependencies and capture relevance between segments of multivariate data, without prior bias based on position~\cite{zerveas2021transformer}.

Formally, at each time point $t$, the output representation is computed via a weighted sum over the value vectors $\bm{z}_{v,t'}^{(i)} \in \mathbb{R}^{\latentdim_v}$, $t'\in [1, \dots, T]$, where the importance weight assigned to the value vector at time $t'$ is computed as a dot-product similarity between its corresponding key vector $\bm{z}_{k,t'}^{(i)} \in \mathbb{R}^{\latentdim_q}$ and a query vector $\bm{z}_{q,t}^{(i)} \in \mathbb{R}^{\latentdim_q}$ at time $t$. 
As a result, given a latent feature $\latentfeat^{(i)}$, a query $\latentfeat_q^{(i)} = [\bm{z}_{q,1}^{(i)}; \dots; \bm{z}_{q,T}^{(i)}]\in \mathbb{R}^{\nooftimepoints \times \latentdim_q}$, a key $\latentfeat_k^{(i)} = [\bm{z}_{k,1}^{(i)}; \dots; \bm{z}_{k,T}^{(i)}] \in \mathbb{R}^{\nooftimepoints \times \latentdim_q}$, and a value $\latentfeat_v^{(i)} = [\bm{z}_{v,1}^{(i)}; \dots; \bm{z}_{v,T}^{(i)}]\in \mathbb{R}^{\nooftimepoints \times \latentdim_v}$ are computed by applying three different trainable affine transformations on $\latentfeat^{(i)}$. The self-attention output for a single attention head (SA) is then computed via a scaled dot-product:
\begin{align}
    \text{SA}(\latentfeat^{(i)}) = \text{softmax}\left(\frac{\latentfeat_q^{(i)} \latentfeat_k^{(i)\top}}{\sqrt{\latentdim_q}}\right)\latentfeat_v^{(i)},
\end{align}
where softmax converts the similarity scores to a probability distribution over the input sequence of length $T$. This operation is performed in parallel for each of the $H$ attention heads (each with its own trainable transformations). The resulting outputs SA$_h \in  \mathbb{R}^{T \times \latentdim_v}$, $h \in [1, \dots, H]$ are first concatenated and finally aggregated into a single representation through a trainable linear transformation $\bm{W}_A \in \mathbb{R}^{H\latentdim_v \times \latentdim}$:
\begin{equation}
\label{equ:self_attention}
    \text{MSA}(\latentfeat^{(i)}) \!\!=\!\! [\text{SA}_1(\latentfeat^{(i)})~\text{SA}_2(\latentfeat^{(i)}) \!\dots\! \text{SA}_H(\latentfeat^{(i)})] \bm{W}_A.
\end{equation}

\subsection{Reconstruction-Based Loss Function}
We aim for the transformer model to extract discriminative latent features that govern the generation of EEG recordings, i.e., to model the input data distribution. To this end, we corrupt each input sample by a novel masking strategy that is uniquely designed for modeling multivariate time-series data such as EEG \cite{zerveas2021transformer}. We train the transformer model via a loss function that minimizes the error between the original (unmasked) recording $\eegimage^{(i)}$ and the corresponding reconstruction $\hat \eegimage^{(i)}$.

Formally, a proportion $\maskratio \in (0,1)$ of each channel $\chnidx \in \{1,\ldots,\noofchn\}$ in each EEG recording $\eegimage^{(i)}$ is dynamically masked at the beginning of each training step by setting the encoder input values at chosen time points to 0. The values at each channel alternate between consecutive masked and unmasked sequences. The number of masked time points follows a geometric distribution with mean $\masklength$, while the number of unmasked time points follows a geometric distribution with mean $\unmasklength=\frac{1-\maskratio}{\maskratio}\masklength$. This transition paradigm is also known as an M/M/1 queue, in which the number of customers in a system is geometrically distributed \cite{gallager2013stochastic}. The resulting masking strategy encourages the transformer to attend on time points preceding and following the masked segments both in individual channels, as well as across the aligned time points in other channels to capture inter-channel dependencies, and has been found more effective than other denoising strategies for downstream tasks, including Bernoulli masking (c.f.~Table \ref{tbl:bernoulli_vs_geometric} \& \cite{zerveas2021transformer}). 

Finally, the reconstruction loss for end-to-end training of our model is the mean-squared reconstruction error. Crucially, the loss is computer over only the set of \emph{masked} time points $\maskedset=\{(\timeidx, \chnidx) \,|\,\text{masked}\,  \,\,\eegimageperelm^{(i)}_{t,m}, \timeidx \in \{1,\ldots,\nooftimepoints\}, \chnidx \in \{1,\ldots,\noofchn\}\}$:
\begin{align}
    \frac{1}{|\maskedset|} \sum_{(\timeidx, \chnidx) \in \maskedset} \,\,\, (\eegimageperelm^{(i)}_{t,m} - \hat \eegimageperelm^{(i)}_{t,m})^2. 
\label{equ:reconstruction_loss}
\end{align}

\subsection{Seizure Identification}
We aim to employ the trained transformer to distinguish between EEG recordings that contain seizures and those which do not; this motivates us to pose unsupervised seizure identification as an anomaly detection problem. Thus, we train the transformer architecture on recordings that \emph{do not} contain seizures. This allows for the learned latent features to capture non-seizure activity \cite{you2020unsupervised}.  
As the transformer is trained to model non-seizure activity, recordings with no seizures are expected to be reconstructed with low error in inference time. In contrast, EEG recordings including seizure activity come from a different distribution, and thus, the model naturally reconstructs such input recordings with a relatively larger error; we use this observation as an indicator for a seizure (c.f.~Section \ref{sec:metrics}). 

We note that the exclusion of seizure recordings from the training set \emph{does not} constitute supervision or require any special annotation, as the default states of patients and healthy individuals alike are non-seizure, whose recordings can be collected and kept separate from the recordings of seizure episodes (which we only use for evaluating our method). In real-life applications, EEG data with no seizure activity can be easily augmented with recordings from healthy individuals, which are trivially accessible compared to the ones from patients experiencing seizures.

\section{Experiments}
\label{sec:experiments}
\subsection{Datasets}
\label{sec:datasets}
We evaluate our method on three publicly available EEG datasets collected at the: (i) Massachusetts Institute of Technology (MIT) and Boston Children's Hospital \cite{shoeb2009application} (ii) University of Pennsylvania (UPenn) and Mayo Clinic \cite{kaggle}, and (iii) Temple University Hospital of  Philadelphia (TUH) \cite{10.3389/fnins.2016.00196}.

The MIT dataset contains EEG recordings acquired on the scalp with $256$ Hz sampling rate from a maximum of $\noofchn=38$ channels. $198$ seizure recordings were labeled w.r.t.~their start and end times. The total duration of non-seizure recordings is $40,800$ seconds and seizure recordings is $2889$ seconds.

The UPenn dataset contains 1-second long EEG recordings acquired intracranially at $500-5000$ Hz from a maximum of $\noofchn=72$ channels. The total duration of non-seizure recordings is $7164$ seconds and seizure recordings is $653$ seconds.

The TUH dataset contains EEG recordings acquired on the scalp with $250$ Hz sampling rate from a maximum of $\noofchn=38$ channels. $1229$ seizure recordings were labeled w.r.t.~their start and end times. The total duration of non-seizure recordings is $49,922$ seconds and seizure recordings is $2600$ seconds.

\subsection{Preprocessing}
\label{sec:preprocessing}
EEG recordings are typically preprocessed to eliminate the powerline noise at 60 Hz \cite{you2020unsupervised}. We first unify the sampling rates in each dataset by downsampling to the smallest sampling rate across all recordings. Then, we filter the recordings via a 4-th order Butterworth bandpass filter with range 0.5-50 Hz. 

To construct samples with the same size, we extract sliding windows over each recording, where each window contains $\nooftimepoints$ time points and overlaps with its consecutive window by 50\%. We choose $\nooftimepoints$ based on the shortest seizure segment in each dataset. In doing so, $\nooftimepoints=1536$ for MIT, $\nooftimepoints=500$ for UPenn, and $\nooftimepoints=462$ for TUH. This process results in $13,600$ windows with non-seizure activity and $963$ windows with seizure activity for MIT, $14,329$ windows with non-seizure activity and $1307$ windows with seizure activity for UPenn, and $54,264$ windows with non-seizure activity and $2826$ windows with seizure activity for TUH. In real-life applications, a minimum seizure window length can be decided by clinical experts, as in UPenn that directly provides 1 second-long seizure recordings.

Moreover, we aim to consistently form $ \nooftimepoints \times \noofchn$ size windows, while not disregarding any channels with potential seizure activity. Thus, to construct samples with the same number of $\noofchn$ channels, we reuse data from other channels for the recordings that have missing data at certain channels, compared to the recording with the largest number of channels in each dataset. Again, in real-life applications, clinical experts can determine which channels to employ or discard for seizure identification. 
Finally, we normalize windows by subtracting the mean and dividing by the standard deviation across all windows to aid the convergence of training \cite{ioffe2015batch}. 

\begin{table*}[t!]
\centering
\setlength{\tabcolsep}{12pt} 
\renewcommand{\arraystretch}{1.5} 
\begin{tabular}{c c c c c c}
\hline
 Dataset & Method & Precision & Recall & Accuracy & AUC\\ 
 \hline
  MIT & \textbf{Unsupervised Transformer} & $\mathbf{0.98 \pm 0.003}$ & $\mathbf{0.9 \pm 0.006}$ & $\mathbf{0.87 \pm 0.006}$ &
  $\mathbf{0.94 \pm 0.023}$\\  
  \cline{2-6}
  & Unsupervised K-means & $0.33 \pm 0.008$ & $0.5 \pm 0.009$ & $0.5 \pm 0.009$ & $0.59 \pm 0.041$\\  
  \cline{2-6}
  & Unsupervised VAE & $0.97 \pm 0.003$ & $0.75 \pm 0.008$ & $0.61 \pm 0.009$ & $0.61 \pm 0.041$\\
  \cline{2-6}
  & Supervised XGBoost & $0.98 \pm 0.003$ & $0.8 \pm 0.007$ & $0.8 \pm 0.007$ & $0.88 \pm 0.031$\\
 \cline{2-6}
 & Supervised ROCKET & $0.98 \pm 0.003$ & $0.74 \pm 0.008$ & $0.78 \pm 0.008$ & $0.86 \pm 0.032$\\
 \cline{2-6}
 & Supervised Transformer & $0.98 \pm 0.003$ & $0.83 \pm 0.007$ & $0.83 \pm 0.007$ & $0.88 \pm 0.031$\\
 \cline{2-6}
 \cline{2-6}
 \cline{2-6}
 \cline{2-6}
 \cline{2-6}
 & Pre-trained 50\% Supervised Transformer & $0.97 \pm 0.003$ & $0.72 \pm 0.008$ & $0.63 \pm 0.009$ & $0.66 \pm 0.021$\\
 \cline{2-6}
 & \textit{Pre-trained 100\% Supervised Transformer} & $\mathit{0.99 \pm 0.002}$ & $\mathit{0.98 \pm 0.003}$ & $\mathit{0.94 \pm 0.005}$ & $\mathit{0.97 \pm 0.017}$\\
 \hline
 UPenn & \textbf{Unsupervised Transformer} & $\mathbf{0.88 \pm 0.01}$ & $\mathbf{0.76 \pm 0.013}$ & $\mathbf{0.68 \pm 0.014}$ & $\mathbf{0.73 \pm 0.027}$\\
 \cline{2-6}
  & Unsupervised K-means & $0.33 \pm 0.014$ & $0.5 \pm 0.015$ & $0.5 \pm 0.015$ & $0.56 \pm 0.028$\\  
  \cline{2-6}
  & Unsupervised VAE & $0.8 \pm 0.012$ & $0.5 \pm 0.015$ & $0.49 \pm 0.015$ & $0.47 \pm 0.027$\\
  \cline{2-6}
  & Supervised XGBoost & $0.87 \pm 0.01$ & $0.62 \pm 0.015$ & $0.6 \pm 0.015$ & $0.65 \pm 0.028$\\
 \cline{2-6}
 & Supervised ROCKET & $0.87 \pm 0.01$ & $0.67 \pm 0.014$ & $0.62 \pm 0.015$ & $0.67 \pm 0.028$\\
 \cline{2-6}
 & Supervised Transformer & $0.87 \pm 0.01$ & $0.69 \pm 0.014$ & $0.62 \pm 0.015$ & $0.64 \pm 0.028$\\
 \cline{2-6}
 \cline{2-6}
 \cline{2-6}
 \cline{2-6}
 \cline{2-6}
 & Pre-trained 50\% Supervised Transformer & $0.86 \pm 0.011$ & $0.77 \pm 0.013$ & $0.63 \pm 0.015$ & $0.64 \pm 0.032$\\
 \cline{2-6}
 & \textit{Pre-trained 100\% Supervised Transformer} & $\mathit{0.92 \pm 0.008}$ & $\mathit{0.85 \pm 0.011}$ & $\mathit{0.82 \pm 0.012}$ &
  $\mathit{0.89 \pm 0.02}$\\ 
 \hline
  TUH & \textbf{Unsupervised Transformer} & $\mathbf{0.92 \pm 0.005}$ & $\mathbf{0.57 \pm 0.009}$ & $\mathbf{0.61 \pm 0.009}$ & $\mathbf{0.57 \pm 0.013}$\\  
  \cline{2-6}
  & Unsupervised K-means & $0.17 \pm 0.007$ & $0.5 \pm 0.009$ & $0.35 \pm 0.008$ & $0.57 \pm 0.013$\\  
  \cline{2-6}
  & \textit{Unsupervised VAE} & $\mathit{0.93 \pm 0.005}$ & $\mathit{0.86 \pm 0.006}$ & $\mathit{0.83 \pm 0.007}$ & $\mathit{0.86 \pm 0.009}$\\
  \cline{2-6}
  & Supervised XGBoost & $0.93 \pm 0.005$ & $0.73 \pm 0.008$ & $0.71 \pm 0.008$ & $0.78 \pm 0.011$\\
 \cline{2-6}
 & Supervised ROCKET & $0.93 \pm 0.005$ & $0.7 \pm 0.008$ & $0.66 \pm 0.008$&$0.74 \pm 0.012$\\
 \cline{2-6}
 & Supervised Transformer & $0.92 \pm 0.005$ & $0.37 \pm 0.009$ & $0.54 \pm 0.009$ & $0.52 \pm 0.012$\\
 \cline{2-6}
 \cline{2-6}
 \cline{2-6} 
 \cline{2-6}
 \cline{2-6}
 & Pre-trained 50\% Supervised Transformer & $0.94\pm 0.005$ & $0.61\pm 0.009$ & $0.75\pm 0.008$ & $0.71\pm 0.025$\\
 \cline{2-6}
 & Pre-trained 100\% Supervised Transformer & $0.93 \pm 0.005$ & $0.66 \pm 0.008$ & $0.7 \pm 0.008$ & $0.72 \pm 0.012$\\
 \hline
\end{tabular}
\vspace{-0.5em}
\caption{Seizure identification performance metrics and confidence intervals on UPenn, MIT and TUH. We compare our transformer-based unsupervised identification method (in bold) with unsupervised methods comprising VAE and t-SNE followed by K-means clustering, as well as supervised methods comprising XGBoost, ROCKET, and the same transformer architecture trained via supervised and pre-trained supervised learning. Best performance for each dataset are in italics.}
\label{tbl:metrics}
\vspace{-1em}
\end{table*}
\begin{figure*}[!t]
\centering
\begin{subfigure}{0.45\textwidth}
\includegraphics[width=0.9\textwidth]{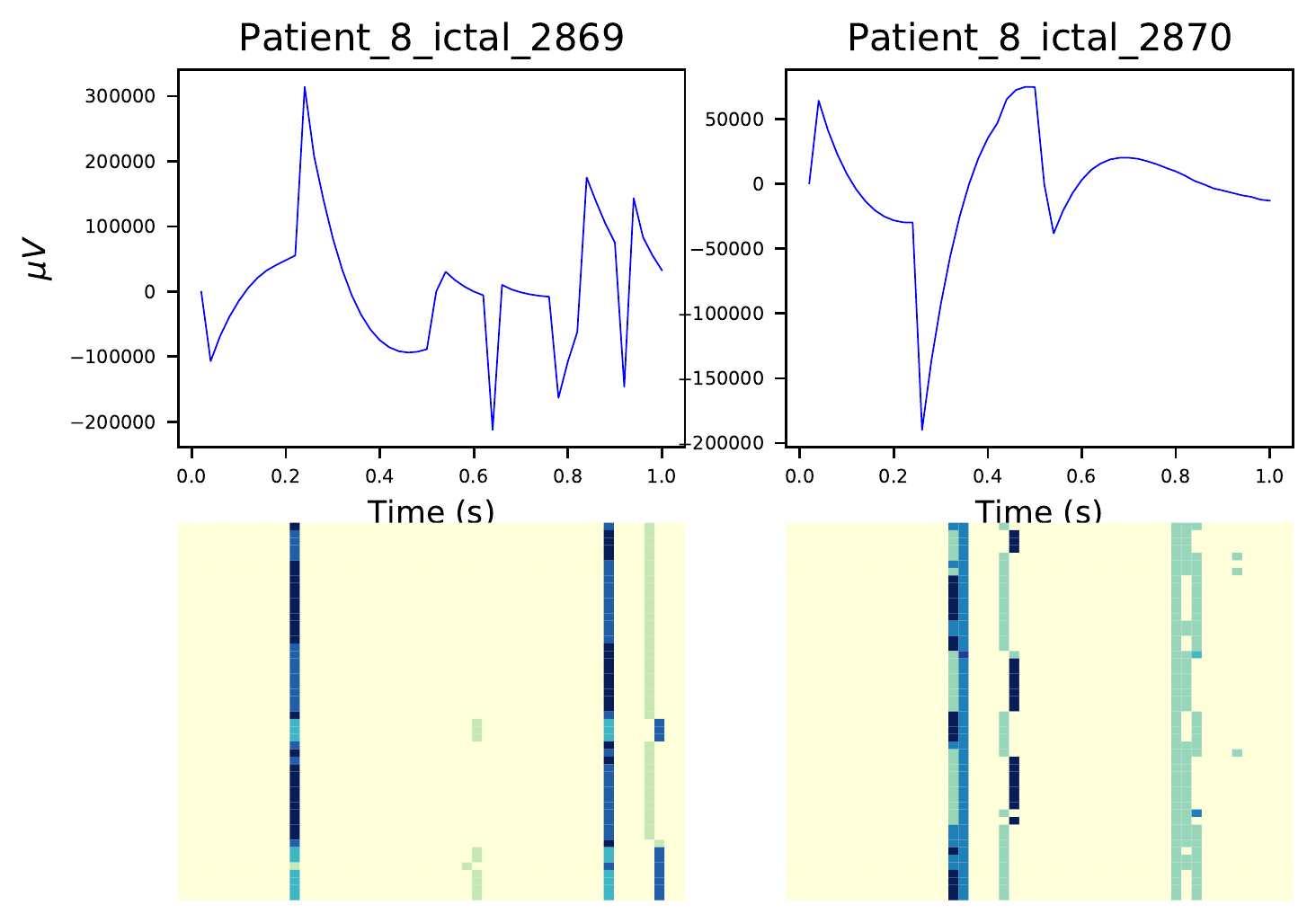}
\caption{Correctly Identified Seizure Windows (True Positive)}
\label{fig:true_pos}
\end{subfigure}%
\hspace{1em}
\begin{subfigure}{0.45\textwidth}
\includegraphics[width=0.9\textwidth]{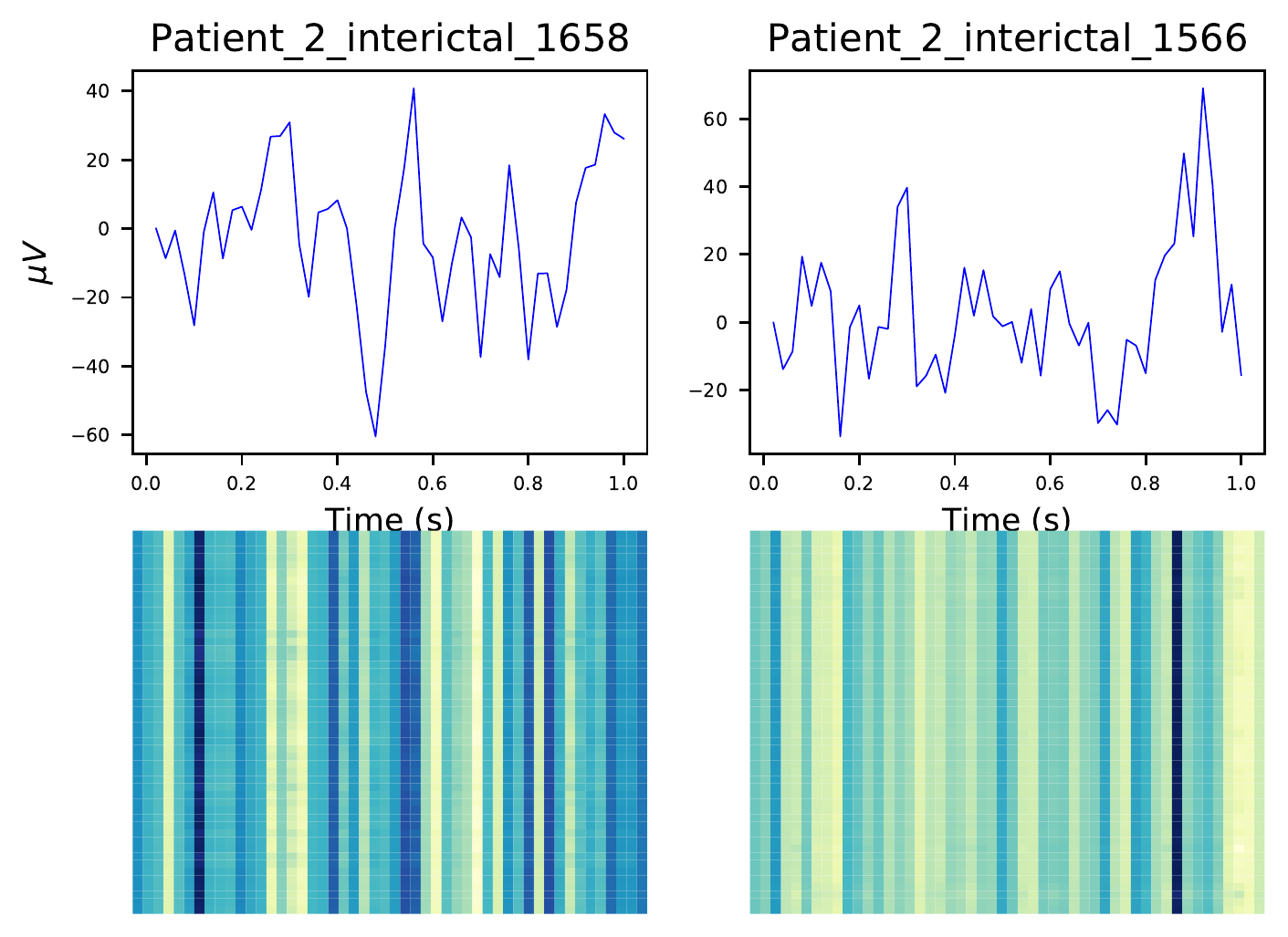}
\caption{Correctly Identified Non-seizure Windows (True Negative)}
\label{fig:true_neg}
\end{subfigure}
\begin{subfigure}{0.45\textwidth}
\includegraphics[width=0.9\textwidth]{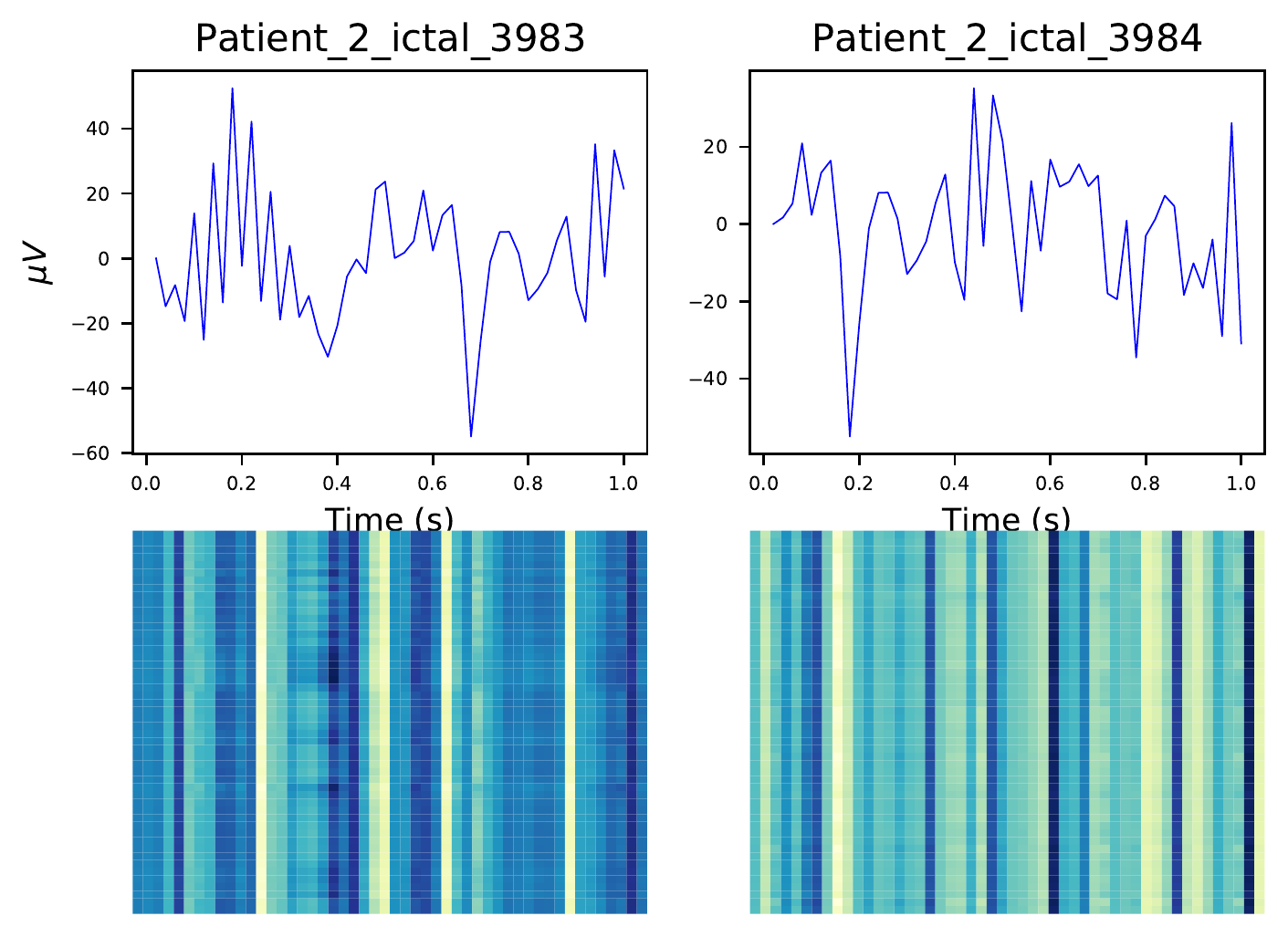}
\caption{Falsely Identified Seizure Windows (False Negative)}
\label{fig:false_neg}
\end{subfigure}%
\hspace{1em}
\begin{subfigure}{0.45\textwidth}
\includegraphics[width=0.9\textwidth]{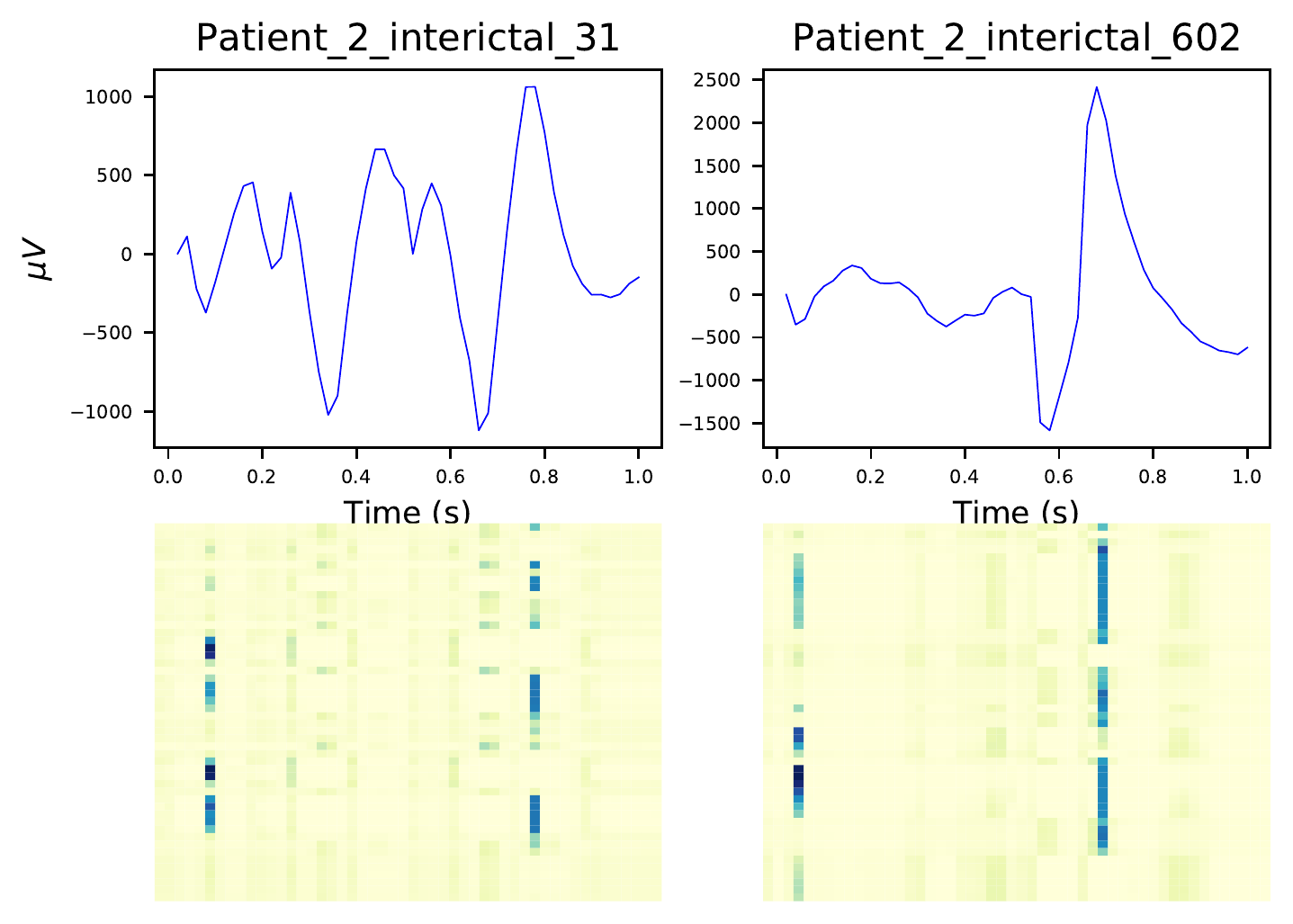}
\caption{Falsely Identified Non-seizure Windows (False Positive)}
\label{fig:false_pos}
\end{subfigure}
\caption{Example EEG windows, corresponding seizure identifications and self-attention weights on UPenn. First and third rows contain example windows of true positive, true negative, false negative, and false positive identifications, respectively. Second and fourth rows contain the corresponding self-attention weight heatmaps computed by the transformer architecture, where darker colors indicate higher importance. For each window, we visualize the channel with the largest reconstruction error.}
\label{fig:seizure_identification_examples}
\vspace{-1em}
\end{figure*}

\subsection{Experiment Setup and Competing Methods}
\label{sec:competing}
We partition all windows containing non-seizure and seizure activity into training, validation, and test sets in a stratified manner, allocating 60\% for training, 20\% for validation, and the remaining 20\% for testing. As baseline methods, we implement shallow and deep learning models for both supervised and unsupervised settings. 

\subsubsection{Unsupervised Learning Methods} 
\label{sec:unsupervised_methods}
For our method, we employ the transformer encoder architecture proposed by Vaswani et al.~(2017), with the modifications of fully-trainable positional encoding, batch normalization and the same hyperparameters suggested by Zerveas et al.~(2021). We train the autoencoder over \emph{only} non-seizure training windows using the unsupervised loss given by Eq.~\eqref{equ:reconstruction_loss}. We monitor the loss value computed over the non-seizure windows in the validation set and use the model that attains the lowest validation loss.

Following the literature on shallow unsupervised methods \cite{8076983}, we reduce the dimension of all EEG windows in the test set to $3$ using the t-Distributed Stochastic Neighbor Embedding (t-SNE) \cite{van2008visualizing} algorithm, and apply K-means clustering \cite{bishop2006pattern} on the resulting windows with two clusters indicating non-seizure and seizure. Moreover, as an unsupervised deep learning baseline, we train a state-of-the-art convolutional VAE \cite{YILDIZ2021106604}.

\subsubsection{Supervised Learning Methods} 
\label{sec:supervised_methods}
First, we employ the same transformer encoder architecture described in Section \ref{sec:architecture} and map the latent features learned from each window to a binary prediction. In doing so, we concatenate all latent features corresponding to all time points of each window into a single vector and apply a fully-connected layer comprising a scalar output with sigmoid activation. We train the resulting architecture via cross-entropy loss over all training windows, employing the same hyperparameters found optimal by Zerveas et al.~(2021). To combat overfitting due to class imbalance in supervised learning, we oversample and augment the seizure windows in training via random reversing and drifting. We monitor the F1-score computed over the validation set and use the model that attains the best validation score.

Moreover, we train state-of-the-art shallow models XGBoost \cite{chen2016xgboost} and ROCKET \cite{dempster2020rocket} over the supervised training set.  XGBoost is a decision-tree classifier using gradient boosting for ensembling. ROCKET transforms time-series using $500$ random convolutional kernels and uses the extracted features to train a ridge regression classifier. Ridge regression hyperparameter is varied in $[10^{-3},10^{3}]$ and best hyperparameter is determined w.r.t.~the accuracy over the validation set.

\subsubsection{Pre-trained Supervised Learning} Finally, we combine the transformer-based seizure identification methods via unsupervised pre-training and supervised fine-tuning \cite{zerveas2021transformer}. Following the unsupervised approach described in Section \ref{sec:unsupervised_methods}, we first pre-train the transformer encoder over non-seizure training windows. Having initialized its weights accordingly, we then fine-tune the model via both non-seizure and seizure training windows, using the same setup described in Section \ref{sec:supervised_methods}. 

\subsection{Evaluation Metrics}
\label{sec:metrics}
To evaluate the seizure identification performance of our approach, as well as the VAE baseline, we use the mean absolute error over the time points and electrode channels in each EEG window from the test set as the corresponding seizure prediction score. For all supervised competing methods, we use the traditional prediction score for inference. 

For all competing methods described in Section \ref{sec:competing}, we report AUC for distinguishing seizure vs.~non-seizure windows in the test set. To compute binary decision metrics, we threshold the prediction score of each window at the value for which the geometric mean of recall and true negative rate is maximal \cite{fawcett2006introduction}. Using the respective threshold, we calculate class-weighted precision and recall, as well as balanced accuracy for binary identification of seizure vs.~non-seizure windows in the test set, considering the imbalanced distribution between the two. In real-life applications, decision thresholds may be determined by clinical experts with respect to the desired trade-off between false positives and negatives \cite{DBLP:journals/corr/abs-1901-03407}.

We report all metrics along with the $95\%$ confidence intervals, which are computed as $1.96 \times \sigma_{A}$, where $\sigma_{A}^2$ is the variance for metric $A$. Variance for AUC is computed by:
\begin{align}
    \sigma_{A}^2\!\!=\!\!\frac{1}{mn}\!\!\left(A(1\!\!-\!\!A)\!\!+\!\!(m\!\!-\!\!1)(P_{x}\!\!-\!\!A^2)\!+\!(n\!\!-\!\!1)(P_{y}\!\!-\!\!A^2)\right),
\end{align}
where $P_{x}=A/(2-A)$, $P_{y}=2A^2/(1+A)$, and $m$, $n$ are the number of seizure and non-seizure windows, respectively \cite{hanley1982meaning}. Variance for other metrics are computed by: 
\begin{align}
    \sigma_{A}^2=A(1-A)/(m+n).
\end{align}

\subsection{Results and Discussion}
\label{sec:results}
\subsubsection{Seizure Identification Performance}
Table \ref{tbl:metrics} shows the seizure identification performance of our transformer-based unsupervised method vs.~supervised and pre-trained supervised transformers, XGBoost, ROCKET, VAE, and t-SNE followed by K-means clustering over all datasets. Our novel transformer-based anomaly detection method establishes a dramatic improvement among all unsupervised methods, by successfully distinguishing between non-seizure vs.~seizure windows with up to \emph{0.94 AUC} and outperforming its state-of-the-art deep learning counterpart VAE by up to $33\%$ AUC on MIT. Clustering on raw EEG windows cannot capture the complex evolution of EEG and predicts all windows as non-seizure. These observations demonstrate the benefit of the transformer architecture for unsupervised anomaly detection in our setting. 

Crucially, despite the lack of seizure labels during training, our unsupervised anomaly detection approach leads to significantly \emph{better seizure identification than all purely supervised learning baselines and the pre-trained transformer fine-tuned with 50\% of the training labels} over UPenn and MIT, by up to $16\%$ recall, $9\%$ accuracy, and $9\%$ AUC. Moreover, unlike supervised learning, class imbalance strongly biases supervised models towards non-seizure predictions and hinders generalization over the distribution of held-out test samples. As a result, unsupervised anomaly detection via transformers establishes a consistently better balance between precision and recall than supervised learning and further demonstrates its benefit in learning from imbalanced datasets such as ours.

The TUH dataset is particularly challenging by being a compilation of several EEG databases collected over years from patients with vast variations in demographic and medical backgrounds \cite{10.3389/fnins.2016.00196}, compared to self-contained UPenn and MIT datasets collected from only 8 and 24 patients, respectively. In this case, our unsupervised transformer still fares significantly better than the purely supervised transformer, while unsupervised VAE outperforms \emph{all} supervised learning baselines, including the pre-trained transformer. These observations further motivate unsupervised learning for our task. 

As expected, the computationally expensive transformer model, which has first undergone unsupervised pre-training and then supervised fine-tuning with \emph{all} training labels, outperforms both purely supervised as well as purely unsupervised transformer models (the latter by a smaller margin). However, our unsupervised anomaly detection method \emph{does not} require ground-truth seizure labels during training as a crucial advantage, while still leading to successful seizure identification. 

\subsubsection{Seizure Identification Examples}
We visualize example EEG windows from UPenn and the corresponding seizure identifications of the unsupervised transformer in the first and third rows of Figure \ref{fig:seizure_identification_examples}, selecting the channel with the largest mean reconstruction error for each window. 
Agreeing with the clinical description of seizures, true seizure windows in Figure \ref{fig:true_pos} contain high-frequency waves with large amplitudes \cite{vespa2019epilepsy}. Meanwhile, true non-seizure windows in Fig.~\ref{fig:true_neg} attain significantly less amplitude changes and spikes compared to true positive windows. 
Note that the seizure patterns cannot be identified w.r.t.~only large amplitude or high frequency, motivating a more sophisticated approach such as ours. For instance, non-seizure windows in Fig.~\ref{fig:false_pos} have a larger amplitude range than the seizure windows in Figure \ref{fig:false_neg}, while the seizure windows in Fig.~\ref{fig:false_neg} contain similar spikes to the non-seizure windows in Figure \ref{fig:true_neg} w.r.t.~amplitude and frequency.

\subsubsection{Benefit of Self-Attention}
We visualize the self-attention weights computed by the last encoder layer of the unsupervised transformer on example EEG windows from UPenn as 2D heatmaps in the second and fourth rows of Figure \ref{fig:seizure_identification_examples}. 
For each time point along the horizontal axis of each heatmap, 
self-attention weights (c.f.~Equation \eqref{equ:self_attention}) from other time points are indicated along the vertical axis. Darker heatmap colors correspond to larger weights and, thus, higher importance. 

It appears that the transformer model within our unsupervised identification method can successfully learn to pay more attention to seizure patterns including high-frequency spikes and waves evolving with large amplitudes \cite{vespa2019epilepsy}. Moreover, when the model predicts the existence of seizures, it shows patterns of focused attention, containing only few time points with large weights (Figures \ref{fig:true_pos} and \ref{fig:false_pos}), while windows identified as non-seizure (Figures \ref{fig:true_neg} and \ref{fig:false_neg}) lead to much more evenly distributed attention. These observations indicate that employing a transformer architecture with self-attention can improve both performance, as well as explainability of seizure identification decisions, by underlining, e.g., spike-wave discharges that are indicative of seizures \cite{vespa2019epilepsy}.

\subsubsection{Effect of Masking Strategy}
Table \ref{tbl:bernoulli_vs_geometric} shows the seizure identification performance of training with our geometric masking strategy against masking each time point independently at random with a Bernoulli distribution. Our approach of unsupervised training with geometric masking consistently leads to better performance than Bernoulli masking, demonstrating its benefit in modeling multivariate data such as EEG.

\begin{table}[t!]
\footnotesize
\centering
\setlength{\tabcolsep}{2pt} 
\renewcommand{\arraystretch}{1.1}
\begin{tabular}{c c c c c c}
\hline
 Dataset & Method & Precision & Recall & Accuracy & AUC\\ 
 \hline
 MIT & Geometric (Ours) & $\mathbf{0.98}$ & $\mathbf{0.9}$ & $\mathbf{0.87}$ & $\mathbf{0.94}$ \\
 \cline{2-6}
 & Bernoulli & 0.98 & 0.85 & 0.85 & 0.9\\
 \hline
UPenn & Geometric (Ours) & $\mathbf{0.88}$ & $\mathbf{0.76}$ & $\mathbf{0.68}$ & $\mathbf{0.73}$\\
 \cline{2-6}
 & Bernoulli & 0.86 & 0.72 & 0.65 & 0.72\\
 \hline
 TUH & Geometric (Ours) & $0.92$ & $\mathbf{0.57}$ & $\mathbf{0.61}$ & $\mathbf{0.57}$\\
 \cline{2-6}
 & Bernoulli & \textbf{0.93} & 0.4 & 0.59 & 0.54\\
 \hline
\end{tabular}
\caption{Effect of masking strategy on seizure identification.}
\label{tbl:bernoulli_vs_geometric}
\vspace{-1.5em}
\end{table}

\section{Conclusion}
We propose a fully-unsupervised transformer-based method for seizure identification on raw EEG. Our method can successfully distinguish between non-seizure and seizure windows and can even achieve significantly better seizure identification performance than state-of-the-art supervised time-series methods, including its purely supervised transformer-based counterpart. Generalizing our method to other applications involving anomalous activity detection on multivariate time-series data is a promising future direction. 

Our unsupervised approach can significantly alleviate the burden on clinical experts regarding laborious and difficult EEG inspections to provide labels indicating segments that contain seizures. Furthermore, if automated identification performance meets clinical requirements, our method can aid availability of seizure diagnoses for the wider public, especially in areas where access to well-trained healthcare professionals is limited.

\bibliographystyle{plain}
\bibliography{mybibfile.bib}

\end{document}